\documentclass[preprint,amssymb,showpacs,supersriptaddress,floatfix]{revtex4-2}
\usepackage{graphicx}
\usepackage{dcolumn}
\usepackage{bm}
\usepackage[utf8]{inputenc}
\usepackage[T1]{fontenc}
\usepackage{mathptmx}
\usepackage{etoolbox}
\usepackage{float}
\usepackage{booktabs}
\usepackage{amsmath,amssymb}
\usepackage[dvipsnames]{xcolor}
\usepackage{subcaption}
\usepackage{mathtools}
\usepackage[normalem]{ulem}

\makeatletter
\def\@email#1#2{
 \endgroup
 \patchcmd{\titleblock@produce}
  {\frontmatter@RRAPformat}
  {\frontmatter@RRAPformat{\produce@RRAP{*#1\href{mailto:#2}{#2}}}\frontmatter@RRAPformat}
  {}{}
}
\makeatother

\usepackage{color}

\usepackage[utf8]{inputenc}
\usepackage[T1]{fontenc}

\begin{document}

\title{Tuning the memristive response of TaO$_x$- based devices  with Ag Nanoparticles}  
\author{R. Leal Martir$^{1,2}$}
\author{A.J.T. van der Ree $^{3,4}$}
\author{M. H. Aguirre $^{5,6,7}$}
\author{G. Palasantzas $^{3,4}$}
\author{D. Rubi$^{8}$}
\author{M. J. Sánchez $^{1,2,*}$}

\affiliation{$^{1}$Instituto de Nanociencia y Nanotecnología (INN),CONICET-CNEA, nodo Bariloche, 8400 San Carlos de Bariloche, Río Negro, Argentina.\\
$^{2}$Centro Atómico Bariloche, Instituto Balseiro (UNCuyo), CONICET, 8400 San Carlos de Bariloche, Río Negro, Argentina.\\ $^{3}$Zernike Institute for Advanced Materials, University of Groningen, 9747 AG Groningen, The Netherlands. \\
$^{4}$CogniGron Center, University of Groningen, 9747 AG Groningen, The Netherlands. \\
$^{5}$Instituto de Nanociencia y Materiales de Aragón (INMA-Unizar), Zaragoza, Spain.\\
$^{6}$Departamento de Física de la Materia Condensada, UNIZAR, Zaragoza, Spain.\\
$^{7}$Laboratorio de Microscopías Avanzadas, UNIZAR-CSIC, Zaragoza, Spain.\\
$^{8}$Laboratorio de Ablación Láser (INN-CONICET-CNEA), Centro Atómico Constituyentes, Gral. Paz 1499, San Martín, Argentina.
}

\email{maria.sanchez@ib.edu.ar}

\begin{abstract}

Defect engineering is a key strategy to control resistive switching (RS) in oxide-based memristive devices, where oxygen vacancy (OV) dynamics governs filament formation and rupture. We investigate the effect of Ag nanoparticles (AgNPs) embedded in the top electrode of Pt/Ta$_2$O$_5$/TaO$_2$/Pt memristors  and analyze their RS behavior and statistical stability. Devices without AgNPs exhibit two hysteresis switching loops  (HSLs) with opposite chiralities, originating from the participation of  the  Pt/Ta$_2$O$_5$ top interface and the Ta$_2$O$_5$/TaO$_2$ bottom interface. Incorporating AgNPs reduces the overall device resistance and selectively suppresses one loop, yielding a single, well-defined switching mode.
Moreover, devices incorporating   AgNPs show markedly reduced cycle-to-cycle variability of the high-resistance state, as confirmed by Weibull analysis, indicating improved endurance and switching reproducibility. Within a filamentary RS framework, we attribute this behavior to local  metallization of the top interface by AgNPs, which partially inhibit OV transport and confines  the RS dynamics to the bottom interface. Numerical simulations with the Oxygen Vacancy Resistive Network (OVRN) model  succesfully reproduce the experimental HSLs, statistical trends, and tunable ON/OFF ratios with AgNPs coverage. 
These findings demonstrate that targeted interface metallization via metallic nanoparticles provides an effective route to control multi-interface RS dynamics and
improve switching stability in without modifying the oxide architecture.

\end{abstract}

\maketitle

\section{INTRODUCTION}
Defect engineering plays a crucial role in the development and optimization of memristive devices, as the presence, distribution, and dynamics of defects critically influence their resistive switching (RS) performance, namely the reversible change in resistance induced by the application of an external electrical stimulus \cite{saw_2008,iel_2016}. By carefully tailoring the concentration and spatial arrangement of defects, it is possible to control the formation and rupture of conductive filaments, tune switching thresholds, and improve device performance in terms of speed, endurance, and retention \cite{baner_2020,kim_2019,tutu_2022,wu_2018}. Moreover, defect engineering enables the modulation of non-volatile memory states and enhances the reproducibility of switching processes, which are essential for reliable integration into neuromorphic computing architectures \cite{liu_2010,baner_2012,choi_2013}. As a result, understanding and manipulating defects at the atomic scale has become a key strategy for advancing memristor technologies toward practical applications.

Within transition-metal-oxide (TMO) memristors, particularly binary oxides such as TiO$_2$, VO$_2$, HfO$_2$, and TaO$_2$ \cite{nandi_2020, ghenzi_2012, son_2008, alvarez_2024}, which exhibit a decrease in local resistivity with decreasing oxygen content \cite{bao_2023, lu_2012, arif_2017}, RS is generally attributed either to the migration of oxygen vacancies (OVs) in regions confined to the oxide/electrode interfaces \cite{sas_2016, Fer_2020, Fer_2024} or to the formation and rupture of OV filaments spanning the oxide layer \cite{saw_2008, waser_2010, bae_2012}. This latter mechanism, known as filamentary resistive switching (FRS), has been extensively investigated using a variety of experimental techniques over the last years \cite{nandi_2020, son_2008, celano_2013, isaev_2023, yang_2012, peng_2012, krishnan_2016, cheng_2021}. Despite its technological promise, FRS remains constrained by reliability challenges, including limited endurance, unstable conductive pathways, and poor data retention.

In multilayer or asymmetric oxide systems, an additional level of complexity arises from the coexistence of multiple (active) interfaces. When two or more interfaces contribute to OV dynamics, distinct switching channels may coexist or compete within the same device, giving rise to multiple hysteresis switching loops (HSLs) with different polarities or chiralities. Such behavior has been reported in TaO$_x$-based systems, where both oxide/electrode and oxide/oxide interfaces can actively participate in RS \cite{Fer_2020, Fer_2024, bae_2012}. While this multi-interface activity provides additional degrees of freedom for tailoring device functionality, it also increases the number of accessible microscopic switching configurations, often resulting in increased cycle-to-cycle variability and reduced endurance. Strategies capable of selectively activating or suppressing specific switching channels, without modifying the oxide stack itself, remain limited.

Defect engineering has therefore been increasingly explored as a means to stabilize filament dynamics and improve device robustness. Strategies such as incorporating Cu nanoparticles into electrodes \cite{liu_2010} or introducing Ag/Cu nanoparticles either within the oxide matrix \cite{Gao_2017, jana_2025} or at electrode interfaces \cite{li_2022} have demonstrated varying degrees of success in enhancing switching uniformity and reliability. However, in many cases the physical mechanisms by which metallic nanoparticles affect the competition between different OV-driven switching channels remain insufficiently understood.

On the theoretical side, most studies restrict the analysis of defect effects to static electric-field simulations \cite{Gao_2017, li_2022, jana_2025} or energy band calculations \cite{Ning_2021, Goul_2022}, largely because the complex dynamics of OV migration under electrical stress is difficult to capture. The Oxygen Vacancy Resistive Network (OVRN) model \cite{LealMartir_2025} overcomes this limitation by explicitly describing (2D) OV dynamics within an array of nanodomains interconnected through a 2D resistor network. This framework enables direct simulations of OV migration and redistribution, naturally incorporates filamentary behavior, and allows the controlled introduction of spatial inhomogeneities, providing a powerful tool to investigate FRS dynamics in complex oxide systems.

In this work, we introduce Ag nanoparticles (AgNPs) into the top electrode of a Pt/Ta$_2$O$_5$/TaO$_2$/Pt memristive device and investigate their impact on RS. Devices without nanoparticles exhibit two hysteresis switching loops (HSLs) with opposite chiralities—clockwise (CW) and counter-clockwise (CCW)—sharing a common low-resistance state. Upon the incorporation of AgNPs, the overall resistance is reduced and the CCW loop is selectively suppressed, yielding a single, well-defined switching mode.

This behavior can be rationalized within a filamentary RS framework. In devices without AgNPs, Schottky barriers at both the Pt/Ta$_2$O$_5$ top interface (TI) \cite{lee_2011, zhuo_2013} and the Ta$_2$O$_5$/TaO$_2$ bottom interface (BI) \cite{Fer_2020} promote OV migration and enable two competing switching channels. Local metallization of the TI induced by the AgNPs partially inhibits OV transport across this interface, effectively confining the RS dynamics to the BI and suppressing one of the channels. As a result, devices with AgNPs exhibit reduced cycle-to-cycle variability of the high-resistance states, as quantified by Weibull analysis, indicating enhanced switching reproducibility and endurance stability. This physical picture is validated using the OVRN model \cite{LealMartir_2025}, which reproduces the experimental HSLs and their statistical trends, and shows that the OV dynamics and ON/OFF ratios can be tuned via the AgNPs coverage.

\section{Methods}

A bilayer TaO$_x$ thin film was grown over a Si/Pt substrate through Pulsed Laser Deposition (PLD) at room temperature. The lower (35 nm) and upper (15 nm) layers were deposited at oxygen pressures of 0.01 mbar and 0.1 mbar respectively. After the film was grown, AgNPs were deposited on top. This was done through plasma sputtering in a home-modified Mantis Nanogen 50 unit \cite{Brink_2014, vanderRee_2024}. The coverage (6$\%$) was measured on a simultaneously NP-deposited TEM grid with a Helios G4 scanning electron microscope at 18 kV using HAADF detector. Although other films were fabricated with higher  coverages and larger average AgNPs diameters, these conditions are not considered in the present study, as they consistently resulted in device failure. Part of the sample was covered during the depositon, leaving a sector of the film without AgNPs to serve as control. The top Pt electrodes were fabricated using a combination of sputtering and optical lithography. Both regions (those with and without NPs) underwent identical processing, resulting in two types of devices on the same film: one incorporating NPs embedded within the Pt contact, and one without NPs.

Scanning electron microscopy with high angular annular dark field detector (STEM-HAADF) were performed using a Cs-probe-corrected Titan 360–300 (ThermoFisher) at a working voltage of 300 kV equipped with a EDS detector AZtec from Oxford Instruments with an estimated error in chemical characterization of $\approx 2 \%$ \cite{Fer_2020}. Lamellae were prepared by Focused Ion Beam performed in a Dual Beam Helios NanoLab 650, using a Ga+ with acceleration voltage from 5kV up to 30kV.
The electrical characterization of the devices was carried out with a Keithley 2636B SMU unit, hooked to a probe station constructed in-house. Device yields were estimated to be approximately 25\% and 33\% for devices without and with nanoparticles, respectively. The electrical characteristics reported below were reproducibly observed across multiple working devices.

\section{Experimental Characterization and electrical response}

\begin{figure}[H] 
\centering
\includegraphics[width=0.9\linewidth]{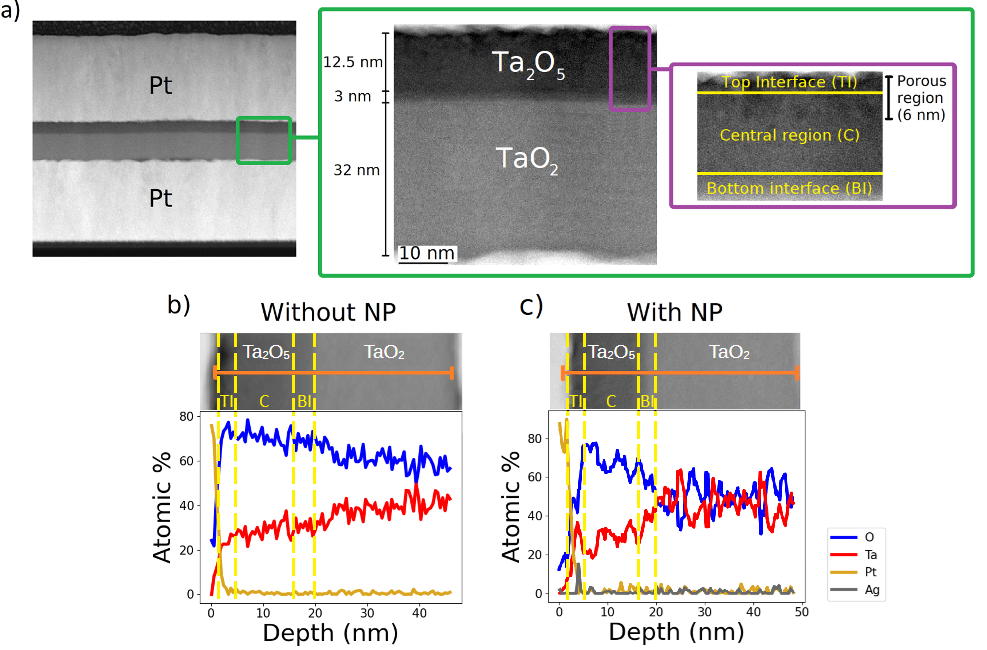}
\caption{a) STEM-HAADF cross-sectional image of a device with Ag nanoparticles (AgNPs) embedded in the top electrode. The active zone of the device togheter with the  top, central and bottom interfaces (TI, C and BI respectively) are zoomed in by the green and violet  rectangles, respectively. See text for details.  EDS linescans of the device taken along the orange line. Ta, O, Pt and Ag species are quantified: b) Without AgNPs, c)  With AgNPs. }
\label{Fig1}
\end{figure}

Figure \ref{Fig1}a) shows a STEM-HAADF cross-sectional image of a device containing NPs in its pristine (virgin) state. The structure of the TaO$_x$ thin film can be divided into three distinct regions: a top layer ($\sim$ 12.5 nm thick, corresponding to Ta$_2$O$_5$), a bottom layer ($\sim$ 32 nm thick, corresponding to TaO$_2$), and a transition region between them ($\sim$ 3 nm thick). The uppermost $\sim$ 6 nm of the film exhibits a porous morphology, in contrast to the denser structure observed in the rest of the layer. For subsequent analysis  we consider  the two interfaces and the central  region scketched in the right panel of FIG. \ref{Fig1}a): the TI, encompassing the Pt/Ta$_2$O$_5$ interface; the BI corresponding to the Ta$_2$O$_5$/TaO$_2$ interface; and  the central  C region containing the Ta$_2$O$_5$. The TaO$_x$ in the film is amorphous, while the AgNPs are crystalline.

The NPs are not visible in the STEM images. However, the presence of NPs was unambigously verified by EDS. Line scans of the device with NPs, shown in FIG. \ref{Fig1}c), reveal a distinct Ag signal near the Pt/TaO$_x$ interface, which is absent in devices without NPs, as seen in FIG. \ref{Fig1}b).  An oxygen concentration gradient is observed throughout both layers of the film, ranging from Ta$_2$O$_5$ near the top electrode to TaO near the bottom electrode.


To investigate the memristive behavior of the system, Hysteresis Switching Loops (HSLs) were measured. In this procedure, voltage pulses (referred to as writing pulses) are applied in a ramp sequence. The remnant resistance of the device is then measured between successive writing pulses by applying a small reading pulse (0.1 V) and plotted as a function of the writing voltage. Further methodological details are provided in Refs.\cite{roz_2010, Fer_2020}.

HSLs and the corresponding I-V curves (not shown) were acquired for devices with and without AgNPs. Both kinds of devices initially exhibited a low resistance state and no electroforming step was necessary. This could be due to the porous structure \cite{chakrabarti_2021} shown in FIG. \ref{Fig1}a), or  to a low density media  that creates high mobility paths \cite{tsuruoka_2015}, which can favor the formation of conductive OV filaments through internal surfaces without the need for an electroforming step.

\begin{figure}[H] 
\centering
\includegraphics[width=0.95\linewidth]{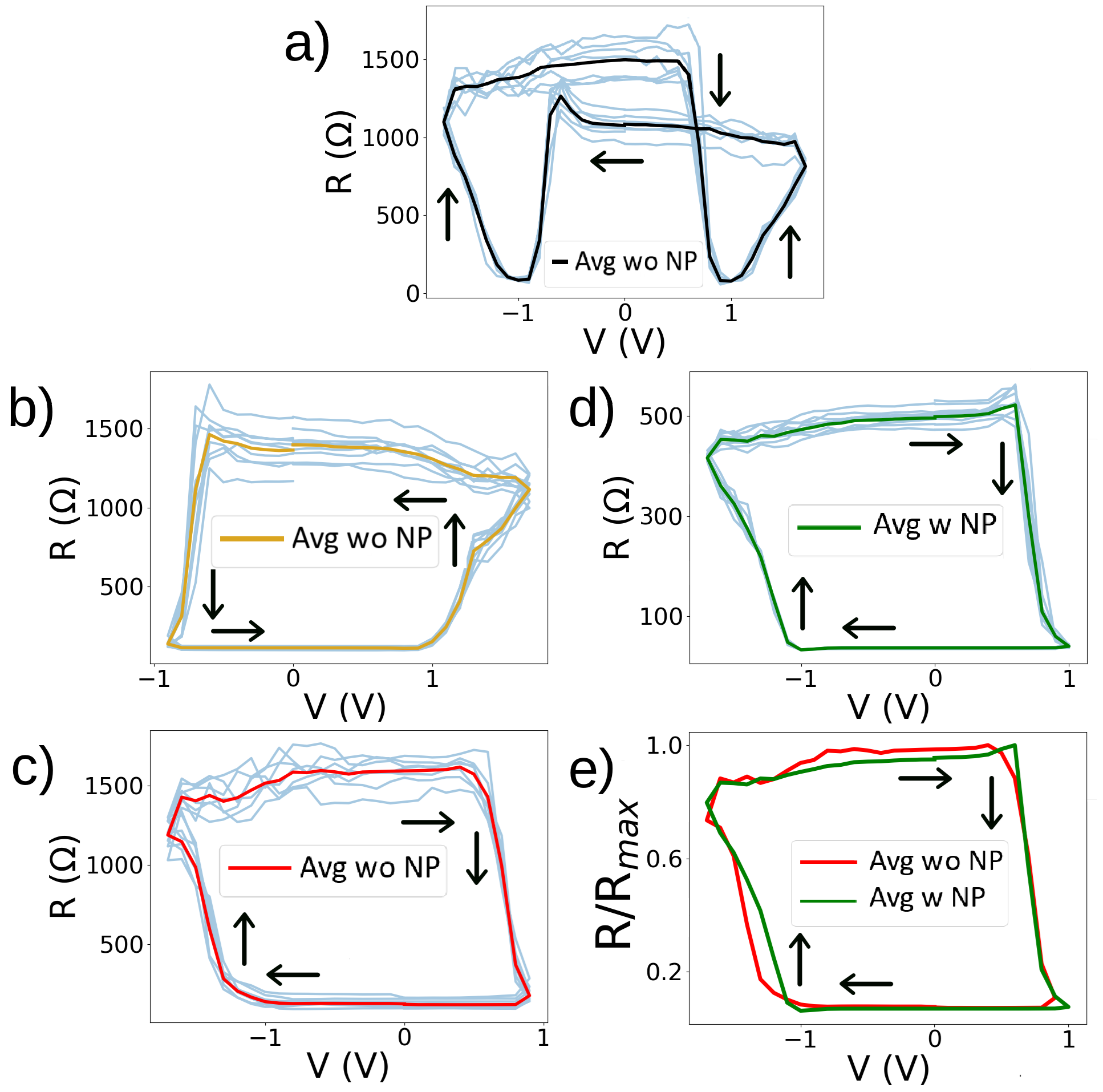}
\caption{Electrical characterization of TaO$_x$ devices with (w) and without (wo) Ag nanoparticles (AgNPs). Measured individual hysteresis switching loops (HSL)  (light blue lines)  and average (Avg) HSL. Arrows indicate the circulation direction. a) Table with legs HSL in a device wo AgNPs. b) Counter-clockwise HSL wo AgNPs. c) Clockwise HSL wo AgNPs. d) Clockwise HSL  in a device w AgNPs. e) Clockwise HSL for devices wo and w AgNPs.}
\label{Fig2}

\end{figure}

In devices without AgNPs, the application of a symmetrical voltage ramp produced a  "table with legs" (TWL) type of  HSL (see FIG. \ref{Fig2}a)). By modifying the symmetry of the electrical stimulus, it was possible to isolate two HSLs with opposite circulations: a CCW loop (FIG. \ref{Fig2}b)) and a CW loop (FIG. \ref{Fig2}c)). The high-resistance  (HR) values were around $\sim$  1.4-1.5k$\Omega$, and slightly different for both, CCW and CW, HSL. On the other hand, the low-resistance (LR) state was essentially the same for both loops and about 120~$\Omega$. Note in addition, that the TWL  also exhibits  two  HR levels, in contrast to the case of symmetric devices for which a single one is obtained \cite{igna_2008, Ge_2014}. In  Sec. \ref{model} we will further elaborate on these results using the OVRN model.

The SET and RESET voltages were similar in magnitude across both loops (0.9~V and 1.5~V, respectively), with their polarity reversing according to the loop circulation. However, SET transitions are observed to be relatively abrupt, which is consistent with a FRS mechanism driven by OV migration.

Devices containing AgNPs exhibited a  markedly different behavior as only the CW loop was experimentally obtained (see FIG.\ref{Fig2}d)). Additionally, the resistance levels were significantly reduced compared to those of devices without NPs, with the HR and LR states  roughly of 450 $\Omega$ and 30 $\Omega$ respectively. Notice that the SET and RESET transitions 
take place at the same voltages and have a similar trend  than in the CW cycle without NPs.
To reinforce these features, the average CW HSLs from both device types were normalized to their maximum  resistance values and plotted together in FIG.\ref{Fig2} e). The near-perfect overlap of these curves suggests that the underlying RS mechanism remains mostly unchanged regardless of the presence of NPs. 

\begin{figure}[H] 
\centering
\includegraphics[width=0.9\linewidth]{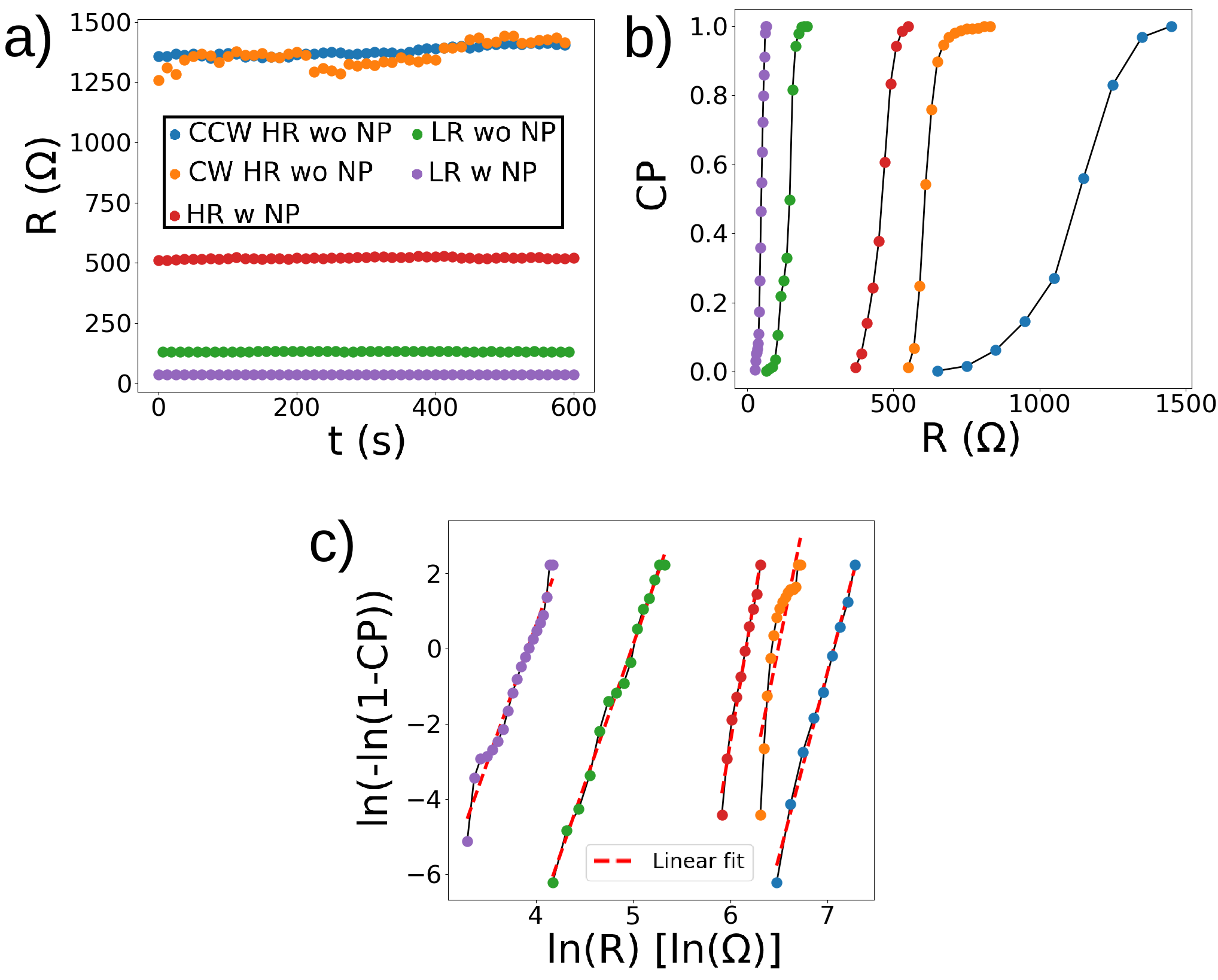}
\caption{ a) Retentivity test showing all resistive levels. b) Cumulative probability distribution function  (CP) obtained as a result of 500 ON/OFF cycles for devices wo and w AgNPs. c) Linear Weibull fittings (dashed lines) of $\ln (-\ln (1- CP))$ vs $\ln(R)$ for each of the resistive levels of devices with and without AgNPs. See text for details.}
\label{Fig3}

\end{figure}

To further investigate the stability and nature of the RS effect, short-term retention tests were performed  and displayed in FIG. \ref{Fig3}a). The resistance states remained stable in both type of devices over  10 minutes, indicating the absence of volatile switching effects regardless of the presence of NPs. To test device durability, 500 ON/OFF cycles were carried out, resulting in a much narrower cumulative probability distribution (CP) for both high and  and low resistance levels in the case of devices  with AgNPs (as seen in  FIG. \ref{Fig3}b)).

The evolution of both the HR and LR states of a given device upon repeated cycling can be analyzed with  the Weibull cumulative probability distribution \cite{Jana2015, Roman_2025},
 given by 
\begin{equation}
    {CP}_{w}(R,\lambda,k) = 1-e^{-(\frac{R}{\lambda})^k},
\end{equation}
with $\lambda$ being the scale parameter, $k$ the shape parameter and R the resistance. 
A typical approach is to use  
    $\ln(-\ln(1-{CP}_{w})) = k \ln(R) - k \ln(\lambda)$,
which displays  linear behaviour  as a function of   $\ln(R)$. In  FIG.\ref{Fig3}c) we used the CP values shown in FIG. \ref{Fig3}b) and  plot $\ln(-\ln(1-CP))$ versus $\ln(R)$. After performing a linear fit, we get the $k$ and $\lambda$ values shown in TAB. \ref{Weibull}.

Incorporation of AgNPs improves the cycle-to-cycle uniformity of the HR states, as evidenced by a steeper cumulative probability distribution and a higher shape parameter $k$, compared to those of the  HR states without AgNPs.  This evidences that devices with AgNPs present reduced statistical dispersion and enhanced switching reproducibility. This improvement is consistent with the AgNPs-induced local metallization of the top interface, which pins OV configurations and effectively reduces the number of accessible microscopic switching pathways. As a result, the RS dynamics becomes predominantly governed by the bottom interface, yielding steeper Weibull cumulative distributions and more stable endurance.

\begin{table}[H]\label{table1}
\begin{center}
\begin{tabular}{lcr}

\toprule
\textbf{Resistive level} & \textbf{k} & \textbf{$\lambda$($\Omega$)}\\		       
\midrule

CCW HR wo AgNP & 9.8 &  7.1\\	

CW HR wo AgNP & 12.8 & 6.4 \\	

LR wo AgNP & 7.4 & 4.9\\

HR w AgNP & 15.4 & 6.2\\

LR w AgNP & 7.3 &  3.9\\

\bottomrule
\end{tabular}\end{center} 
	\caption{Shape parameter k and scale parameter $\lambda$ values obtained via Weibull fittings of the resistive levels of devices w and wo AgNPs.}
\label{Weibull}
\end{table}

\section{MODELLING}\label{model}

The described electrical response  can be explained and modelled in terms of  FRS.  In this scenario, the LR state  is associated to  formed OV filaments, while their retraction and regrowth, mainly at the TI and BI-driven by external applied voltage biases of opposite polarities, give rise to the LR $\leftrightarrow$ HR transitions and, therefore, to the RS effect.

\begin{figure}[H] 
\centering
\includegraphics[width=0.8\linewidth]{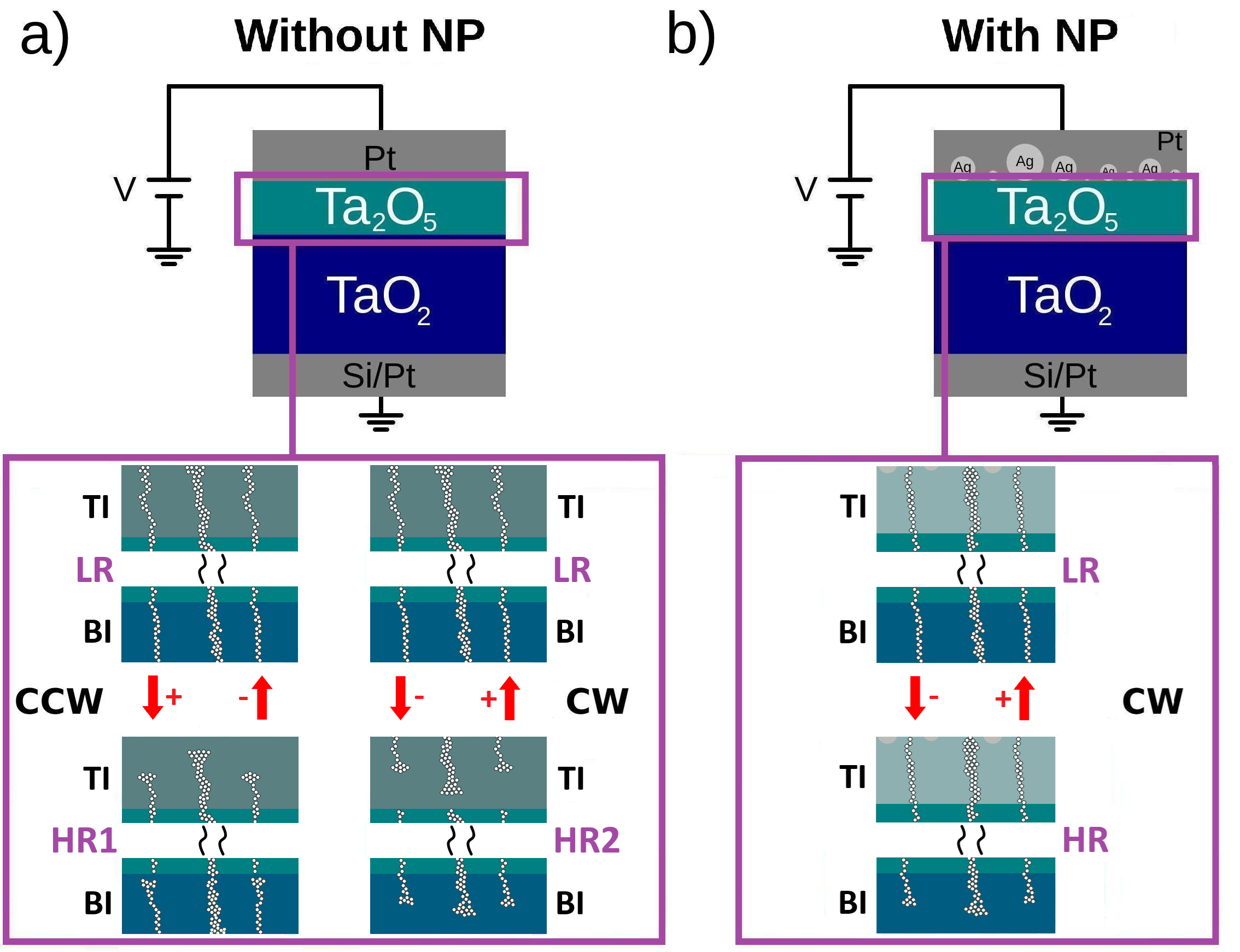}
\caption{Scheme of the device with the active region for RS- comprising the TI, C and BI -enclosed by the violet rectangle. a) Without NPs:OV filaments  bridge the Ta$_2$O$_5$ layer (LR state). When an external stimulus is applied, these filaments retract/grow at the TI/BI interfaces (HR1 and HR2 states). b) With AgNPs: RS is mostly restricted to the BI, giving a single CW cycle. See text for details.}
\label{fig4}
\end{figure}

In the device without NPs, and depending on  the history of the applied voltage, the system can reach three distinct resistive states: a LR state and two high resistance states (HR1 and HR2). The first corresponds to OV filaments fully formed. Meanwhile, the latter ones are related to OV partially expelled from the TI (HR1) and the BI (HR2), with the concomitant break up of the previously formed OV filaments. FIG. \ref{fig4}a) shows a sketch of these processes, that we describe in more detail in what follows:
Considering the LR state as the initial state,  a positive voltage applied to the top electrode drifts  OV (being positively charged defects) towards the bottom electrode. As OV migration takes place predominantly near the interface regions where the Schottky barriers enhace the local electric field, OV filaments are thus expelled from the TI towards the C region, where the OV dynamic is mostly suppressed. 
In addition, due to the lower resistivity of the BI, the electric field intensity is attenuated there compared to that in the TI (as discussed below), so OV filaments break to a lesser extent or only partially. Altogheter this behavior  leads to the HR1 state.
Afterwards,  a negative bias applied to the TE can re-build the OV filaments, leading back to the LR state. This switching between LR and HR1 defines the CCW cycle depicted in FIG. \ref{fig4}a) left.

On the other hand, starting from the same LR  state, a negative voltage drives OV toward the top electrode  and from the BI towards the C region. This process leads to the HR2 shown in FIG. \ref{fig4}a) right. In analogy with the previous description, a subsequent positive bias  will re-build or enalarge  OV filaments at both interfaces, recovering the LR state. This hole process produces the CW cycle between LR and HR2. 

For the present analysis we assume that  the LR state is shared by  the two cycles—CCW (HR1$\mathrel{\mathop{\rightleftarrows}}$LR) and CW (HR2$\mathrel{\mathop{\rightleftarrows}}$LR)—allowing access to either HR1 or HR2 depending on the polarity of the applied stimulus. This assumption  relies on  the experimental measurements that give the same LR states  for both CW and CCW HSL (see FIG. \ref{Fig2} b) and c)). In spite of the fact that our numerical simulations support this hypothesis (see below), it would be ideal to have  experimental images of OV distributions, which are not accessible with current  capabilities.

With the introduction of AgNPs at the top electrode, the TI is metallized in the vicinity of the NPs. This creates low resistivity regions that anchor OV dynamics due to the weak electric field, inhibiting  OVs to be expelled towards the C region, thus rendering the HR1 state inaccesible. OV dynamics is (mostly) limited to the BI region (see FIG. \ref{fig4}b)), concomitantly with  a single  CW HSL, as was obtained in the experiment (FIG. \ref{Fig2} d)).

To validate the proposed response, we  perform simulations using the OVRN model \cite{LealMartir_2025}, which accounts for resistance variations in a resistor network by considering the migration of OV  under the influence of an external electrical stimulus. Owing to the dependence of the oxide’s local resistivity on its specific oxygen stoichiometry, the model enables a precise description of  the evolution of the overall device resistance as OV  dynamics takes place. In addition, the 2d nature of the model allows to simulate the filamentary behaviour, successfully reproducing features such as the abnormal SET process observed in TaOx-based devices \cite{park_2015}.

The OVRN considers the oxide as a N × M  grid of nanodomains in the x-y plane, each one labelled with an index k (l $\leq$ k $\leq$ N × M). The resistivity ($\rho_k$) of a nanodomain is determined by its local OV concentration ($\delta_k$) according to 

\begin{equation}
    \rho_{k} = \rho_M - \rho_m \tanh{[A(\delta_k - \delta_M)]},
    \label{resis}
\end{equation}

where $\rho_m = (H\rho - L\rho)/2$ and $\rho_M = (H\rho + L\rho)/2$, with L$\rho$ and H$\rho$ being the maximum and minimum resistivites attainable by a given domain respectively. 
In the simulations we have considered different values of $\rho_m$  and $\rho_M $ for the TI, C and BI  respectively, to take into account the different resistivity ranges of each region (See Table \ref{Params} for  the parameters' values. The parameter $\delta_M$ denotes a threshold vacancy concentration, such that for $\delta_k$ = $\delta_M$ the resistivity of the domain k is $\rho_M$. The constant A controls the sharpness of the transition between the high (and low) resistivity states. Since OV act as n-type dopants, their accumulation locally reduces the resistivity. The hyperbolic tangent function effectively captures this behavior, while not allowing the resistivity to decrease indefinitely. 
The first N$_{TI}$  rows were assigned to the TI region, the last N$_{BI}$ rows to the BI region and the remaining sites to the central (C) region.
Each simulation step at time $t$ starts with a defined OV profile $\delta_k (t)$, which determines a resistivity profile $\rho_k (t)$. An external voltage $V(t)$ is applied to the top electrode (the bottom electrode remains grounded). Using Kirchhoff’s laws, the local voltage drops $\Delta V_k^{l}$ ($l = x$ or $l = y$) across each site are computed along with the total resistance 
$R(t)$ of the network. The  OV migration rates (transition probability per unit time) from a domain k to its four nearest neightbours k' are obtained through
\begin{equation}
    p(k,k')= \delta_k(t) (1-\delta_{k'}(t)) \exp{(-V_{\alpha}^{l}+\Delta V_k^{l})},
\end{equation}
where $V_{\alpha}^l$ is the activation energy for OV diffusion, expressed in units of the thermal energy $kT$. The vacancy profile at time $t+\Delta t$  is then updated  to
\begin{equation}
    \delta_k (t + \Delta t) = \delta_k (t) + \Sigma_{k'} (p(k',k) - p(k,k')) \Delta t,
\end{equation}
and with these values the new resistivity profile $\rho_k(t+\Delta t)$  is obtained. 

Using the OVRN model, two sets of simulations were performed, without  and with AgNPs, respectively. All simulations were initialized with a randomized OV profile and  in accordance with experimental observations, the system was set  initially in a low resistance state, corresponding to fully formed OV filaments. A voltage ramp was applied until the HSL stabilized. We consider  100  random generated initial OV configurations and  the  HSL  is obtained by averaging over all the realizations.

\begin{figure}[H] 
\centering
\includegraphics[width=1\linewidth]{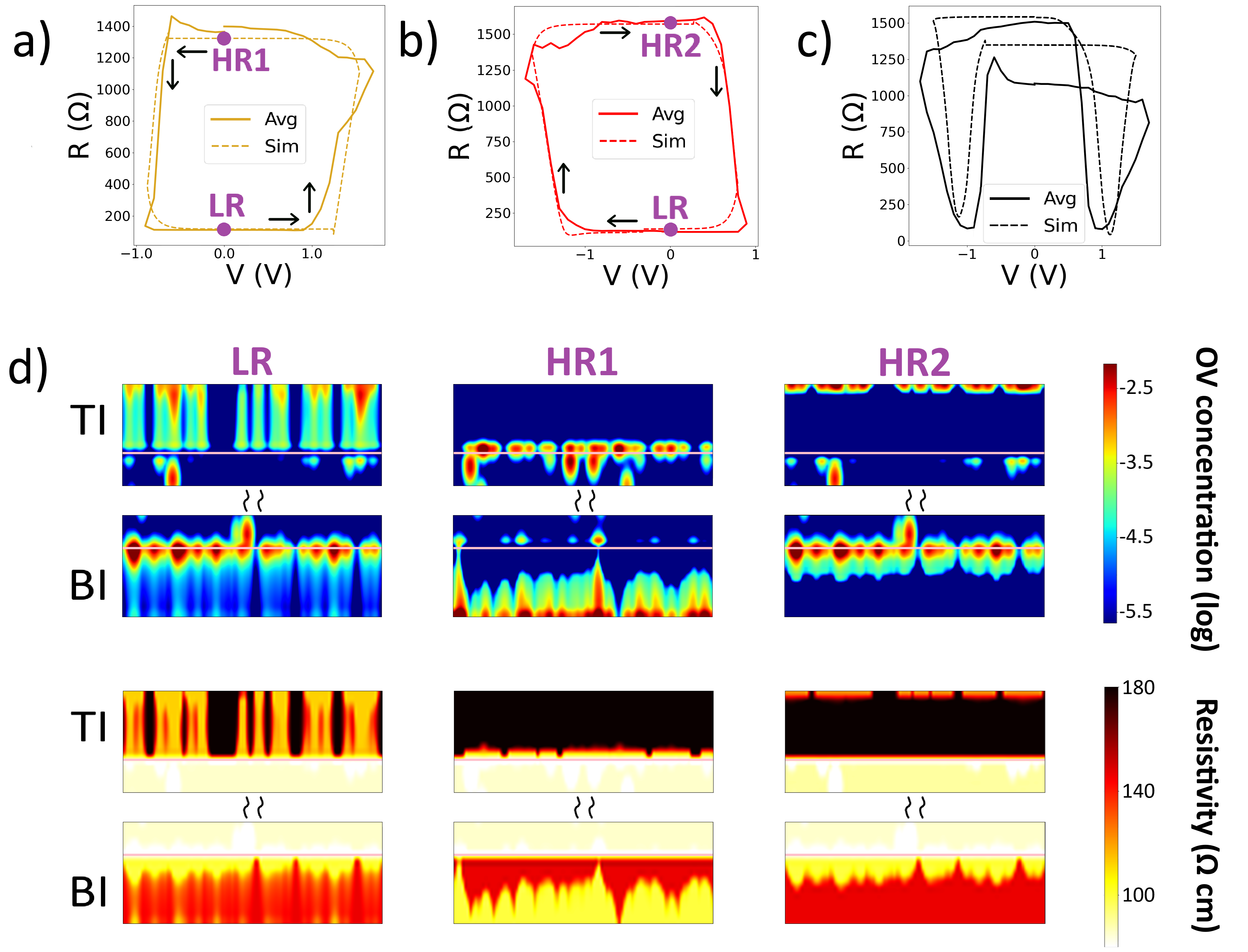}
\caption{ Top panels: Average experimental (solid lines) and simulated (dashed line) HSLs  for the device without NPs. a) TWL, b) CCW   and c) CW, HSLs. d) OV concentration (upper panel) and resistivity  profiles (bottom panel) near the TI and BI,  for the LR (left panels), HR1 (central panels) and HR2 (right panels) resistance states indicated in a) and b).}
\label{fig5}
\end{figure}

In the simulations without NPs, the TI and BI  were chosen more resistive than the C region, consistent with the formation of Schottky barriers at both interfaces \cite{Fer_2020,lee_2011,zhuo_2013}. Simulated CCW (FIG. \ref{fig5}a)) and CW (FIG. \ref{fig5}b)) cycles have been properly isolated by stimulating the system with a non-symetrical voltage ramp. The average experimental HSL were  also plot  for comparison, showing an excellent  agreement with the simulated ones. Likewise, for  a symmetrical voltage ramp, the  simulated  HSL displays a  TWL (FIG. \ref{fig5}c)), in quite good agreement with the experimental one. It is worth noting, however, that certain features of the experimentally observed high-resistance states are not reproduced by the simulations. These include, for example, the resistance of the CCW level in the TWL configuration and the upward drift in resistance observed between the RESET and SET processes. Such discrepancies may arise from the presence of slight diffusive processes linked to memristive filamentary mechanisms, triggered by the presence of high oxygen vacancy gradients \cite{Zhang_2022} which are not fully collected by the model.

The OV and resistivity profiles correspondent to the high resistance states, HR1 and HR2, in addition  to  those of the LR state, are displayed in FIG. \ref{fig5}d). Depending on the polarity of the applied voltage, the OV filaments retract/grow at both the TI and BI. Note that, due to the weak electric field in the C region, the OV distribution there remains largely unchanged; therefore, we show only the OV and resistivity profiles of the C region near  the TI and BI, respectively.

\begin{table}[H]
\begin{center}
\begin{tabular}{lcr}

\toprule
\textbf{Parameter} & \textbf{Numerical value (a.u.)} & \textbf{Value in units }\\		       
\midrule

Applied voltage  V (abs. value) & 0 to 1600 & 0 V to 1.6 V \\	

A & 10$^7$ & 200 nm$^{2}$ \\	

$\delta_M$ & 1.4 $\cdot$ 10$^{-4}$ & 0.3 nm$^{2}$ \\	

KT@RT (300 K) & 1 & 25 meV\\
V$_{\alpha}$ (in units of KT) & 13 & 13 \\	

N & 100 & 60 $\mu$m \\	

M & 100 & 15 nm \\	

M$^{TI}$ & 10 & 1.5 nm \\	

M$^{C}$ & 80 & 12 nm \\	

M$^{BI}$& 10 & 1.5 nm \\	

$\rho_M^{TI}$ -  $\rho_m^{TI}$& 77 - 32 & 5.6 $\cdot$ 10$^{2}$ $\Omega \cdot $m - 2.8 $\cdot$ 10$^{2}$ $\Omega \cdot $m\\

$\rho_M^{C}$ -  $\rho_m^{C}$ & 28 - 5 & 2 $\cdot$ 10$^{2}$ $\Omega \cdot $m - 0.1 $\cdot$ 10$^{2}$ $\Omega \cdot $m\ \\

$\rho_M^{BI}$ -  $\rho_m^{BI}$ & 60 - 22 & 4.3 $\cdot$ 10$^{2}$ $\Omega \cdot $m - 1.6 $\cdot$ 10$^{2}$ $\Omega \cdot $m \\

$\rho_M^{NP}$ -  $\rho_m^{NP}$ & 41 - 24 & 2.9 $\cdot$ 10$^{2}$ $\Omega \cdot $m - 1.7 $\cdot$ 10$^{2}$ $\Omega \cdot $m \\

\bottomrule
\end{tabular}\end{center} 
\caption{Parameters used in the OVRN model converted from a.u to physical units. See text for more details.}
\label{Params}
\end{table}

An interesting outcome that captures the experimetal results, is   the emergence of distinct   HR1 and HR2 values  arising, as we have  already mentioned, from  the  different resistivity ranges assumed for  the TI and BI . In particular, the  resistivity of the BI was taken lower  than the TI one -in consistency with the higher metallic content obtained by the EDS scans reported in FIG. \ref{Fig1}b)–c).
Notice that even for  the LR state, the OV  concentrations  and resistivities profiles are quite different in both interfaces  and this affects the subsequent OV dynamics.
When  a positive stimulus is applied, the attained HR1 state still presents several complete filaments in the BI, while the OV filaments have almost dissapeared from  the TI. In contrast, in the HR2 state both interfaces exhibit broke up filaments and, as a consequence, the  HR2 value  results slighty higher than the HR1 one. In addition, we also considered different  Schottky barriers for both interfaces. As the BI is formed between two stoichiometries of the same oxide, it is expected to have a lower Schottky barrier than the  Pt/Ta$_2$O$_5$ interface. This  reduces the local electric field and the OV migration along the BI, in consistency with the described behaviour.

The predominance of the interfaces in controlling OV migration due to the strong electric fields there developed, also stabilizes the electric response, producing a more reliable RS effect than that associated with filament rupture due to Joule heating \cite{sas_2016, zhu_2017}. 

\begin{figure}[H] 
\centering
\includegraphics[width=0.9\linewidth]{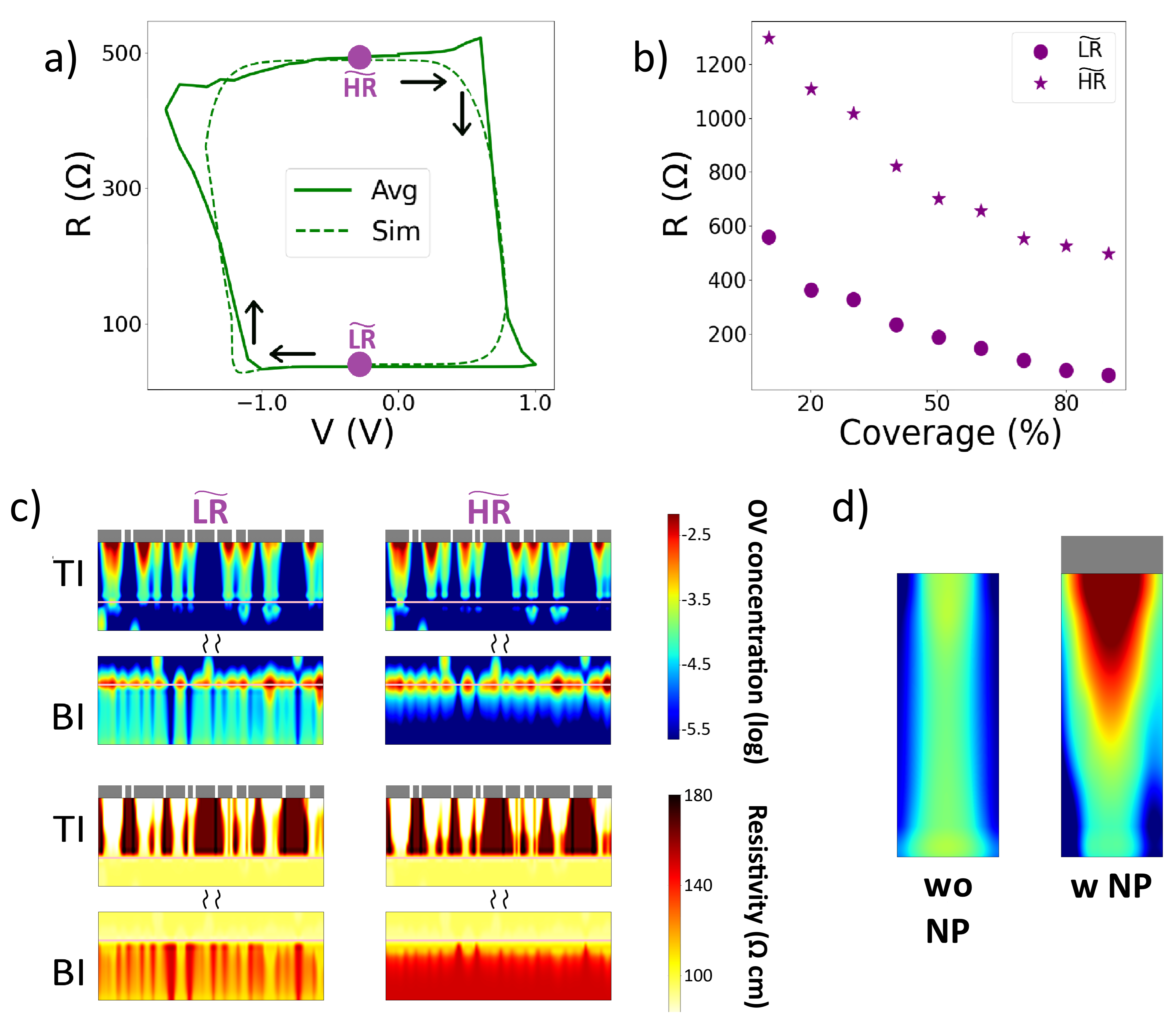}
\caption{a) Average experimental (solid line) and simulated (dashed lined) HSL for the system with AgNPs. b) Simulated $\widetilde {HR}$  and $\widetilde{LR}$  levels as a function of the AgNP coverage. Notice that the simulated HSL in a) corresponds to  a 80 $\%$ of coverage. c) Simulated OV concentration (upper panel) and resistivity  profiles near the TI and BI (bottom panel) for the $\widetilde{LR}$ (left) and $\widetilde{HR}$  (right)  states indicated in a). The AgNPs positions are scketched by the grey regions above the TI. d)  OV filament without (left) and with AgNPs (right).}
\label{fig6}
\end{figure}

The effect of incorporating nanoparticles was successfully reproduced using the OVRN model, as is shown by  the CW  HSL depicted in FIG. \ref{fig6}a). This behaviour was captured in the simulations by assuming that the TI is locally metallized by the presence of AgNPs. These low resistivity regions have been considered  in the simulations by changing   properly in Eq. (\ref{resis}) $\rho_M,\rho_m\rightarrow\rho_M^{NP}, \rho_m^{NP}$, in those network sites where NP have been incorporated. Therefore, AgNps anchor OVs generating persistently a low-resistance state in the TI. Consequently, essentially the BI remains active in the process as is illustrated  in FIG. \ref{fig6}c), and therefore the HSL obtained is the CW one.
Additionally,  the partial short-circuiting of the TI  due to the AgNPs  lowers the resistance of this region concomitant with the lowering of the  overall device resistance, and in agreement with the experimental observations.

The gradual incorporation of AgNPs was simulated and its effect in the overall device resistance  shown in FIG. \ref{fig6}b). An 80\% coverage in the 2D network simulation, which roughly translates to the 6\% measured coverage (see Suppl. Mat. for more details) yields the values measured in the experimental system. 

It is interesting to note that the AgNPs have a noticeable effect in OV filament shape, as shown in FIG. \ref{fig6}d). Filaments in the TI of devices with out (wo) NPs tend to be cylindrical in shape, owing to the fact that the electric field caused by the applied electrical stimulus is always strong enough to move the OV along the interface. The anchoring of OVs caused by the AgNPs gives conical shaped filaments, with more OV concentrated near the AgNPs.

\section{ DISCUSSION AND CONCLUSIONS} 

In conclusion, we have demonstrated that the incorporation of Ag nanoparticles (AgNPs) into the top electrode of Pt/Ta$_2$O$_5$/TaO$_2$/Pt memristive devices provides an effective means to tailor their resistive switching (RS) behavior. Devices without nanoparticles exhibit two distinct hysteresis switching loops (HSLs) with opposite chiralities, which originate from the participation of two active interfaces—Pt/Ta$_2$O$_5$ and Ta$_2$O$_5$/TaO$_2$—and share a common low-resistance state. Introducing AgNPs leads to a significant reduction of the overall device resistance and selectively suppresses one of the switching loops, resulting in a single, well-defined switching mode.

This behavior can be consistently understood within a filamentary RS framework. In the absence of AgNPs, Schottky barriers at both interfaces enhance the local electric field and promote oxygen vacancy (OV) migration, enabling two competing switching channels associated with filament retraction and regrowth at either interface. When AgNPs are incorporated, the top interface becomes locally metallized, partially inhibiting OV transport across this region and effectively confining the RS dynamics to the bottom interface. As a consequence, only one switching channel remains active.

 Although the forming-free behavior of the devices and the presence of AgNPs could, in principle, suggest an electrochemical metallization scenario \cite{tsuruoka_2015,Muk_2025}, several observations support an OV-driven filamentary mechanism instead. In particular, forming-free switching was also observed in devices without NPs, no evidence of Ag incorporation within the oxide layer was detected (See Suppl. Mat.), and the interface-selective switching behavior is inconsistent with the formation of metallic Ag filaments.

Beyond the qualitative modification of the switching characteristics, AgNPs incorporation also leads to a marked improvement in endurance stability. Weibull analysis of the cumulative resistance distributions reveals a reduced cycle-to-cycle variability of the high-resistance states in devices with AgNPs compared to their nanoparticle-free counterparts, indicating enhanced switching reproducibility. This increased stability can be attributed to the reduced number of accessible microscopic switching configurations when OV dynamics is confined to a single active interface.

The proposed physical picture is supported by numerical simulations based on the Oxygen Vacancy Resistive Network (OVRN) model, which successfully reproduce the experimental HSLs both with and without AgNPs, as well as the observed reduction in statistical dispersion. The simulations further show that OV filaments predominantly grow and retract at the active interfaces, remaining largely static in the central region, and that the RS characteristics and ON/OFF ratios can be continuously tuned by varying the AgNPs coverage.

 More broadly, our findings highlight the potential of NP incorporation as a general strategy to tune memristive behavior across a wide range of transition-metal oxides. Because the mechanism identified here relies on controlling OV dynamics at complementary interfaces, similar approaches can be extended to systems in which multi-interface effects govern resistive switching. The incorporation of metallic nanoparticles can, thus, be used to define or suppress specific switching channels, enabling deliberate control over device performance.

\section*{SUPPLEMENTARY MATERIAL}

See the supplementary material for details on the STEM-HAADF and EDS study of  the sample post-electrical stimulation, 
 together with details on the OVRN model and the calculations performed to estimate the AgNPs coverage in the numerical simulations.

\section*{ACKNOWLEDGMENTS}
The authors would like to thank the ‘Laboratorio de Microscopías Avanzadas’  at ‘Universidad de Zaragoza’ for providing access to their instruments. We acknowledge the support from EU-H2020-RISE project MELON (Grant No. 872631) and  ANPCyT (Grant Nos.
PICT2019-0654, PICT2019-02781, and PICT2020A-00415).

\bibliography{reference}

\end{document}